# A Lasso-OLS Hybrid Approach to Covariate Selection and Average Treatment Effect Estimation for Clustered RCTs Using Design-Based Methods

## May 2020

Peter Z. Schochet, Ph.D.
Senior Fellow
Mathematica Policy Research, Inc.
P.O. Box 2393
Princeton, NJ 08543-2393
Phone: (609) 936-2783
Fax: (609) 799-0005
pschochet@mathematica-mpr.com

## Abstract

Statistical power is often a concern for clustered RCTs due to variance inflation from design effects and the high cost of adding study clusters (such as hospitals, schools, or communities). While covariate pre-specification is the preferred approach for improving power to estimate regression-adjusted average treatment effects (ATEs), further precision gains can be achieved through covariate selection once primary outcomes have been collected. This article uses design-based methods underlying clustered RCTs to develop a Lasso-OLS hybrid procedure for the post-hoc selection of covariates and ATE estimation that avoids model overfitting and lack of transparency. In the first stage, lasso estimation is conducted using cluster-level averages, where asymptotic normality is proved using a new central limit theorem for finite population regression estimators. In the second stage, ATEs and design-based standard errors are estimated using weighted least squares with the first stage lasso covariates. This nonparametric approach applies to continuous, binary, and discrete outcomes. Simulation results indicate that Type 1 errors of the second stage ATE estimates are near nominal values and standard errors are near true ones, although somewhat conservative with small samples. The method is demonstrated using data from a large, federally funded clustered RCT testing the effects of school-based programs promoting behavioral health.

## 1. Introduction

In randomized control trials (RCTs) of interventions, random assignment is often performed at the group level (such as a hospital, school, or community) rather than at the individual level. A common challenge for these clustered designs is obtaining large enough sample sizes to detect target average treatment effects (ATEs) within study resources. Statistical power is often a concern due to the typical high cost of including study clusters and the need to inflate standard errors of the estimated ATEs to account for the correlation of outcomes between individuals in the same cluster.[1,2,3,4]

Estimating ATEs using regression models that control for baseline covariates is an effective and commonly used approach for increasing precision for clustered RCTs.[4,5,6] For these designs, the use of covariates, such as baseline measures of the primary outcomes, can increase precision by explaining the variation in outcomes between and within clusters. Accordingly, covariate selection is an important design issue for clustered RCTs, especially in the typical case where data are collected and analyzed at the individual level, so that many candidate covariates could be available for analysis.

One way to address the covariate selection problem is to pre-specify covariates in RCT registries and design documents prior to data analysis, for example, using prior information about the strength of outcome-covariate relationships. This approach has been recommended by authors across a range of fields.[7,8,9,10,11] Pre-specification has the advantage that it yields impact estimates with nominal Type 1 errors in repeated samplings, is replicable and transparent, and avoids the criticism that covariates were selected to obtain favorable findings.

Major RCT registries across disciplines, however, do not mandate that covariates be pre-specified, such as in Medicine (ClinicalTrials.gov), Education (sreereg.icpsr.umich.edu),

Economics (socialscienceregistry.org), and Open Science Registries (osf.io/registries). Similarly, major evidence review clearinghouses—such as the Cochrane Collaboration in Health, the What Works Clearinghouse in Education, and the Clearinghouse for Labor Evaluation and Research in Economics—do not require pre-specification of covariates for RCTs to meet evidence standards. Accordingly, many RCT evaluations do not pre-specify covariates (or specify general categories only), instead selecting specific covariates once primary outcomes have been measured.

An advantage of this data-driven selection approach is that it can improve precision by identifying strong, unanticipated outcome-covariate relationships. However, it can suffer from the criticism that model covariates and their transformations were selected to yield favorable findings. Further, statistical inference under this approach must overcome potential model overfitting if many candidate covariates are available for analysis, thereby biasing standard error estimates downward.

This article discusses a data-driven approach for selecting amongst candidate covariates for clustered RCTs that maintains key advantages of the pre-specification approach. The approach involves pre-specifying a fully replicable *process* for selecting covariates rather than pre-specifying the covariates themselves. Our sequential approach decouples covariate selection from impact estimation to increase transparency, similar in spirit to the approach of Tsiatis et al.[12] The first stage of the approach involves the use of Least Absolute Shrinkage and Selection Operator (lasso) procedures for the greedy selection of covariates[13,14,15], which is adapted to clustered RCTs and finite population design-based ATE ratio estimators that underlie experiments[16,17]. We prove under regularity conditions that as the number of clusters, $m$, approaches infinity, our lasso estimator is $\sqrt{m}$-consistent for parameter estimation and can achieve consistent model selection, thereby extending the seminal work of Knight and Fu[18] to clustered RCTs. These asymptotic results rely on a new finite population central limit theorem (CLT) for weighted least squares (WLS) estimators under our RCT working model.

The second stage of the approach involves ATE estimation, where the selected covariates from the first stage are applied to the design-based WLS estimator. This two-stage estimator is thus a clustered version of the Lasso-OLS hybrid estimator found in the literature for non-clustered settings.[19,20,21,22] The approach is nonparametric in that it makes no assumptions about the distribution of potential outcomes; applies to continuous, binary, and discrete outcomes; and allows for general weighting schemes. We conduct simulations showing that the two-stage impact estimator achieves nominal Type 1 errors with standard errors close to the true ones. We also conduct an empirical analysis using data from a large-scale clustered RCT, co-funded by two federal agencies, that tested the effects of school-based interventions promoting social and character development, to demonstrate how the procedure can improve precision without sacrificing technical rigor.

We establish the asymptotic properties of the first-stage lasso estimator assuming a fixed number of covariates, $v$, and not for the more general case where $v = v_m$ grows with $m$. We adopt this classic framework, because we consider RCT settings where baseline covariates are obtained from relatively short surveys or from administrative records data with fixed record lengths, and not, for example, from gene expression or document or image classification settings where $v$ is typically very large (that is, when $v \gg m$). Further, we do not formally examine the statistical properties of the second stage ATE estimator, where we instead use simulations to examine its statistical performance. Belloni and Chernozhukov[21], Belloni et al.[22], and Zhang and Zhang[23] discuss statistical properties of Lasso-OLS hybrid estimators for non-clustered RCTs.

This article contributes to the vast and growing literature on the lasso in several ways, where we highlight the literature closest to our setting. First, we extend the literature on design-based lasso estimation developed for non-clustered designs to the clustered setting. Bloniarz et al.[24] discuss asymptotic properties of design-based lasso adjustments for non-clustered RCTs (with $v \gg m$), where they specify separate lasso models for the treatment and control groups, which are then combined to yield their ATE estimator. We do not adopt the approach of estimating separate models by treatment status due to resulting degrees of freedom losses that can materially reduce power for clustered RCTs that often contain relatively few clusters (and our focus is not on ATE estimation for baseline subgroups). Rather, we consider a pooled lasso model across the two research groups, where the model includes a treatment status indicator and a common set of covariates. In this setting, lasso is unlikely to select the treatment status indicator, which motivates our Lasso-OLS hybrid approach for ATE estimation. Another related article is McConville et al.[25] who consider design-based lasso survey regression estimators of population totals from finite population sampling, but do not consider clustered RCTs or ATE estimation.

A second contribution of this article is that we consider lasso estimation for clustered designs using a design-based framework rather than a model-based framework found in the literature. For instance, Bondell et al.[26], Muller et al.[27], and Belloni et al.[28] consider a linear mixed model lasso framework for clustered RCTs to simultaneously select covariates and estimate ATEs. In contrast, our approach decouples covariate selection and ATE estimation using a design-based framework. Relatedly, there is a literature examining the asymptotic properties of penalized regression models for correlated data using the generalized estimating equations (GEE) framework, although these studies have not focused directly on ATE estimation.[29]

The rest of the article proceeds as follows. Section 2 discusses design-based estimators for clustered RCTs using a potential outcomes framework. Section 3 discusses lasso estimation and Section 4 discusses our Lasso-OLS hybrid ATE estimator. Section 5 presents simulation results and Section 6 presents empirical analysis findings. Section 7 provides conclusions.

## 2. Design-Based Estimators for Clustered RCTs

Following Schochet et al.[17], we consider an RCT design with $m$ clusters where $m^1 = mp$ clusters are randomized to the treatment group and $m^0 = m(1-p)$ are randomized to the control group, where $p$ is the treatment group assignment rate ($0 < p < 1$). We assume a sample of $n_j$ individuals in cluster $j$, with $n$ total individuals. We assume baseline and outcome data are collected at the individual level. We also assume the stable unit treatment value assumption (SUTVA), applied to clustered designs, where an individual's potential outcomes depend only on the individual's cluster treatment assignment and not on the treatment assignments of other clusters in the sample (although the outcomes of individuals within the same cluster could be correlated).[30] A second SUTVA condition is that a study unit offered a particular treatment cannot receive different forms of the treatment.

Let $Y_{ij}(1)$ be the potential outcome for individual $i$ in the treatment condition and let $Y_{ij}(0)$ be the potential outcome for the same individual in the control condition. Potential outcomes can be continuous, binary, or discrete. Let $T_j$ equal 1 for clusters randomly assigned to the treatment condition and 0 for control clusters. Because of randomization, we have that $T_j \perp \left( Y_{ij}(1), Y_{ij}(0) \right)$ for all $i$ and $j$.

We assume a finite population model, where potential outcomes are assumed fixed for the evaluation, so that study results pertain to the study sample but not more broadly. This scenario is realistic when sites, clusters, and individuals are purposively selected for the study (for example, study volunteers), which often occurs in practice.

The ATE parameter of interest for the finite population model for the clustered design is

$$\beta_1 = \frac{\sum_{j=1}^m w_j(\bar{Y}_j(1) - \bar{Y}_j(0))}{\sum_{j=1}^m w_j} = \bar{\bar{Y}}(1) - \bar{\bar{Y}}(0), \tag{1}$$

where $\bar{Y}_j(1) = \frac{1}{w_j}\sum_{i=1}^{n_j} w_{ij}Y_{ij}(1)$ and $\bar{Y}_j(0) = \frac{1}{w_j}\sum_{i=1}^{n_j} w_{ij}Y_{ij}(0)$ are mean cluster-level potential outcomes in the treatment and control conditions, respectively, $w_{ij} > 0$ is the individual-level weight, and $w_j = \sum_{i=1}^{n_j} w_{ij}$ is the cluster-level weight. Depending on the research questions of interest, the weights can be set, for example, so that intervention effects pertain to the average individual ($w_{ij} = 1$ and $w_j = n_j$) or the average cluster ($w_{ij} = 1/n_j$ and $w_j = 1$). The weights can also help adjust for survey nonresponse bias or other design factors.

Under SUTVA, the data generating process for the observed outcome, $y_{ij}$, is

$$y_{ij} = T_j Y_{ij}(1) + (1 - T_j)Y_{ij}(0). \tag{2}$$

This relation states that we can observe $y_{ij} = Y_{ij}(1)$ for those in the treatment group and $y_{ij} = Y_{ij}(0)$ for those in the control group, but not both. This model generalizes, to clustered RCTs, the design-based model for non-clustered designs using a potential outcomes framework.[31,32,33,34]

Rearranging (2) generates the following regression model:

$$y_{ij} = \beta_0 + \beta_1(T_j - p^*) + u_{ij}, \tag{3}$$

where $\beta_1 = \bar{\bar{Y}}(1) - \bar{\bar{Y}}(0)$ is the ATE parameter, $p^* = \frac{1}{\sum_{j=1}^m w_j}\sum_{j=1}^m T_j w_j$ is the weighted treatment group sampling rate, $\beta_0 = p^*\bar{\bar{Y}}(1) + (1 - p^*)\bar{\bar{Y}}(0)$ is the mean potential outcome, and the

"error" term, $u_{ij}$, can be expressed as

$$u_{ij} = T_j(Y_{ij}(1) - \bar{\bar{Y}}(1)) + (1 - T_j)(Y_{ij}(0) - \bar{\bar{Y}}(0)).$$

Unlike the usual regression model for correlated data, in our setting $u_{ij}$ is random solely because $T_j$ is random (see Freedman[35], Imbens and Rubin[33], and Lin[36] for a discussion of this approach for non-clustered RCTs and Yang and Tsiatis[37] for super-population versions). Further, the model does not satisfy key assumptions of the usual regression model for correlated data, because, over the randomization distribution, $u_{ij}$ does not have mean zero, $u_{ij}$ is heteroscedastic, $Cov(u_{ij}, u_{ij'})$ is not constant for individuals in the same cluster, $Cov(u_{ij}, u_{i'j'})$ is nonzero for individuals in different clusters, and $u_{ij}$ is correlated with the regressor $(T_j - p^*)$.[16] Note that this framework is nonparametric because it makes no assumptions about the distribution of potential outcomes, and allows treatment effects to differ across individuals and clusters.

Consider weighted least squares (WLS) estimation of a variant of (3), using the weights, $w_{ij}$, where the explanatory variables include a $1 x k$ vector of fixed, centered baseline covariates, $\tilde{\mathbf{x}}_{ij} = (\mathbf{x}_{ij} - \bar{\bar{\mathbf{x}}})$, with associated parameter vector $\boldsymbol{\gamma}$, where $\bar{\bar{\mathbf{x}}} = \frac{1}{\sum_{j=1}^m w_j} \sum_{j=1}^m \sum_{i=1}^{n_j} w_{ij} \mathbf{x}_{ij}$. The covariates, unaffected by the treatment, can be at the individual or cluster level. We assume the covariate matrix has full rank, and for reasons discussed below, that $m^1 - kp^* - 1 > 0$ and $m^0 - k(1 - p^*) - 1 > 0$ so there are sufficient degrees of freedom for variance estimation.

Note that in the design-based framework, the covariates do not enter the true RCT model in (3) and the ATE estimand does not change, but the covariates will increase precision to the extent they are correlated with the potential outcomes. Further, we do not need to assume for the working model that the true conditional distribution of $y_{ij}$ given $\mathbf{x}_{ij}$ is linear in $\mathbf{x}_{ij}$. We do not consider models that interact $\tilde{\mathbf{x}}_{ij}$ and $\tilde{T}_j$ due to associated degrees of freedom losses that can seriously reduce the power of clustered RCTs that, in practice, often contain relatively few clusters for cost reasons.

Schochet et al.[17] (Theorem 1) prove that under SUTVA, randomization, and some regularity conditions (discussed in the next section), the WLS estimator for the ATE parameter under our RCT working model with covariates, $\hat{\beta}_1$, is consistent and asymptotically normal as $m$ approaches infinity for a hypothetical increasing sequence of finite populations. In principle, parameters should be subscripted by $m$, but we hereafter omit this notation for simplicity. Schochet et al.[17] also show that a consistent (upper bound), plug-in variance estimator based on regression residuals averaged to the cluster level that performs well in simulations is

$$\hat{\text{Var}}\left(\hat{\beta}_1\right) = \frac{1}{(1-R^2)}\left(\frac{s^2(1)}{m^1} + \frac{s^2(0)}{m^0}\right), \qquad (4)$$

where

$$s^2(1) = \frac{1}{(m^1 - kp^* - 1)(\bar{w}^1)^2} \sum_{j:T_j=1}^{m^1} w_j^2 \left(\bar{y}_j - \hat{\beta}_0 - (1-p^*)\hat{\beta}_1 - \tilde{\bar{\mathbf{x}}}_j\hat{\boldsymbol{\gamma}}\right)^2;$$

$$s^2(0) = \frac{1}{(m^0 - k(1-p^*) - 1)(\bar{w}^0)^2} \sum_{j:T_j=0}^{m^0} w_j^2 \left(\bar{y}_j - \hat{\beta}_0 + p^*\hat{\beta}_1 - \tilde{\bar{\mathbf{x}}}_j\hat{\boldsymbol{\gamma}}\right)^2;$$

$R^2$ is the R-squared value from a regression of $\tilde{T}_j$ on the covariates; $\bar{w}^1 = \frac{1}{m^1}\sum_{j:T_j=1}^{m} w_j$ and $\bar{w}^0 = \frac{1}{m^0}\sum_{j:T_j=0}^{m} w_j$ are mean cluster-level weights; $\bar{y}_j = \frac{1}{w_j}\sum_{i=1}^{n_j} w_{ij}y_{ij}$ are cluster-level mean outcomes; $\tilde{\bar{\mathbf{x}}}_j = \bar{\mathbf{x}}_j - \bar{\bar{\mathbf{x}}}$ are centered cluster-level covariates with $\bar{\mathbf{x}}_j = \frac{1}{w_j}\sum_{i=1}^{n_j} w_{ij}\mathbf{x}_{ij}$; and $\hat{\beta}_0$, $\hat{\beta}_1$, and $\hat{\boldsymbol{\gamma}}$ are parameter estimates. Hypothesis testing can be conducted using t-tests with $(m - k - 2)$ degrees of freedom, which is based on the number of clusters, not individuals. While degrees of freedom losses from models with individual level covariates are typically less than $v$, using $v$ yields conservative Type 1 errors, so that is the approach used here.[17,38]

We note a few features of (4). First, variances differ across the two research groups because we allow for heterogeneous treatment effects. Second, the same variance estimator results using non-centered data in the regressions instead of centered data. Third, the covariates will only affect the ATE estimates and increase precision if mean covariate values vary across clusters. Fourth, the estimator pertains to binary, discrete and continuous outcomes. Finally, a variant of (4) subtracts out the term, $\frac{1}{(m-1)}\sum_{j=1}^{m}\frac{w_j^2}{\bar{w}^2}[Y_{ij}(1) - \bar{\bar{Y}}(1) - (Y_{ij}(0) - \bar{\bar{Y}}(0))]^2$, a bound on the heterogeneity of treatment effects across clusters that enters the asymptotic variance formula.[17,39] However, this variant has been shown to perform worse than (4) in simulations, so is not used.[17]

The variance estimator in (4) is based on regression residuals *averaged to the cluster level*, with separate additive terms for the treatment and control groups. Intuitively, the model is estimated using individual-level data, but standard errors are calculated using residual sums of squares based on cluster-level residuals. This suggests an alternative estimation approach— relevant to our lasso approach presented in the next section—where WLS is run using data averaged to the cluster level. Variance estimators using cluster- or individual-level data will be identical for models with cluster-level covariates only, so there is no efficiency loss from data aggregation in these cases. For models with individual-level covariates, efficiency losses using cluster-level averages are tolerable if $k$ is reasonably small.[40]

## 3. Lasso Estimation for Clustered RCTs

Suppose $v$ candidate baseline covariates are available for analysis. In this section, we discuss lasso covariate selection procedures that align with the objective function of the design-based estimator discussed above, are replicable and decoupled from ATE estimation for transparency, minimize overfitting, and exhibit asymptotic properties found in the literature.

### 3.1. Lasso model

Lasso for covariate selection is a commonly used penalized regression approach that selects covariates by shrinking some regression coefficients to zero.[13,14,15,19] Lasso, which was developed for non-clustered settings, estimates coefficients by minimizing a least squares objective subject to the constraint that the sum of the absolute values of the model coefficients is bounded above by some positive number. Lasso has the property that its regularization leads to sparse solutions, which is suitable for clustered RCTs that often contain limited numbers of degrees of freedom.

In our context, lasso can be adapted to the design-based framework for clustered RCTs by minimizing the following penalized loss function using data *averaged to the cluster level*, pooled across the treatment and control groups:

$$\widehat{\boldsymbol{\delta}}_{lasso} = \mathrm{argmin}_{\boldsymbol{\delta}} \left\{ \sum_{j=1}^{m} \frac{w_j}{\overline{w}} (\widetilde{\overline{y}}_j - \beta_1 \widetilde{T}_j - \widetilde{\overline{\mathbf{x}}}_j \boldsymbol{\gamma})^2 + \lambda (|\beta_1| + \sum_{q=1}^{v} |\gamma_q|) \right\}, \qquad (5)$$

where $\widetilde{\overline{y}}_j = \overline{y}_j - \overline{\overline{y}}$ and $\widetilde{\overline{\mathbf{x}}}_j = \overline{\mathbf{x}}_j - \overline{\overline{\mathbf{x}}}$ are centered cluster-level mean outcomes and covariates, standardized to have standard deviation 1 (so no intercept is included); $\boldsymbol{\delta} = (\beta_1 \ \boldsymbol{\gamma}')'$ is the parameter vector; and $\lambda = \lambda_m$ is the tuning parameter that controls the amount of regularization (shrinking). As $\lambda$ increases, more shrinking occurs and more parameter estimates are set to 0. Note that in the design-based framework, lasso identifies covariates for the linear working model only, and not necessarily for the true model linking outcomes and covariates as is assumed for model-based approaches.

We include $\tilde{T}_j$ in (5) to conform to our design-based framework, even though it is unlikely to be selected in many RCT settings. This is because treatment effects are typically very small relative to the predictive power of baseline covariates; this motivates our Lasso-OLS estimator. Thus, our approach will typically select covariates under the null hypothesis of zero ATEs.

Note that the objective function in (5) only selects covariates based on the strength of their relationships with the outcome and does not penalize covariates based on their associations with $\tilde{T}_j$. This is because in RCTs, selecting covariates that are correlated with $\tilde{T}_j$ but not with the outcome yields upwardly biased standard error estimates.[7,41,42] Belloni et al.[22] discuss estimating separate lasso models to predict treatment selection and outcomes, and then taking the union of the two selected covariate sets to estimate treatment effects. However, their focus is on settings with uncertainty about which variables and transformations are important confounds. In our RCT setting, however, the treatment assignment mechanism is fully determined by randomization, and covariates do not enter the true data generating process underlying experiments, so there are no confounds (we do not address missing data or study attrition that can cause confounds).

The lasso model in (5) could be estimated using individual-level data instead of cluster-level data. However, while this approach could increase precision, it runs the risk of identifying covariates that primarily explain outcome variation across individuals within clusters that do not improve precision under the design-based framework (see (4) above). Thus, we use cluster-level data to ensure lasso identifies covariates that explain between-cluster variation.

Cross-validation (CV) can be used to select $\lambda$.[43] A common choice to balance the bias-variance tradeoff is 5-fold (or 10-fold) CV where the data are partitioned into 5 (or 10) random groups for training and validation.[14,44,45] In our setting, a complication with this approach is that different random partitions of the sample could yield different covariate selections, which could impede replication. There also might not be sufficient degrees of freedom available for data partitioning for designs with small numbers of clusters. To address these issues, we adopt a leave-one-out CV approach, which is fully replicable and applicable to designs with small cluster samples. It yields lower bias but higher variance than higher-fold CV.[14]

### 3.2. Asymptotic distribution of the lasso estimator

In this section, we provide a result (proved in Appendix A) on the asymptotic distribution ($\sqrt{m}$-consistency) of lasso parameter estimation for (5) for fixed $v$ that generalizes findings from Knight and Fu[18] to our clustered RCT setting. We assume that the lasso model in (5) is sparse in the sense that the working model contains both non-zero and zero parameters, although it is not known which parameters are in which group. We assume that as $m \to \infty$, the treatment assignment rate, $p$, is (approximately) constant so that the numbers of treated and control clusters also grow: $m^1 \to \infty$ and $m^0 \to \infty$.

Our asymptotic result relies heavily on a new finite population CLT for parameter estimates from our working regression model using cluster-level averages. The CLT builds on Theorem 1 in Schochet et al.[17] who develop a finite population CLT for ATE estimation but not for covariate estimation under our framework. The conditions underlying the CLT (and hence for lasso estimation) require some definitions. First for $t \in \{1,0\}$, we define the residualized potential outcomes as $\widetilde{\widetilde{\varepsilon}}_j(t) = \widetilde{\widetilde{Y}}_j(t) - \widetilde{\widetilde{\mathbf{x}}}_j \boldsymbol{\gamma}$, where $\widetilde{\widetilde{Y}}_j(t) = \bar{Y}_j(t) - \bar{\bar{Y}}(t)$. We also define the vector of covariate-residual cross-products as $\bar{\mathbf{r}}_j(t) = \widetilde{\widetilde{\mathbf{x}}}_j' \widetilde{\widetilde{\varepsilon}}_j(t)$ and the vector of centered cross-products as $\widetilde{\widetilde{\mathbf{r}}}_j(t) = \bar{\mathbf{r}}_j(t) - \bar{\bar{\mathbf{r}}}(t)$, where $\bar{\bar{\mathbf{r}}}(t) = \frac{1}{\sum_{j=1}^{m} w_j} \sum_{j=1}^{m} w_j \bar{\mathbf{r}}_j(t)$. Further, we define the following variances and covariances of $\widetilde{\widetilde{\varepsilon}}_j(t)$ and $\widetilde{\widetilde{\mathbf{r}}}_j(t)$ and the weights, covariates, and potential outcomes:

$$S_\varepsilon^2(t) = \frac{1}{m-1} \sum_{j=1}^m \frac{w_j^2}{\overline{w}^2} \widetilde{\widetilde{\varepsilon}}_j(t)^2, \qquad \mathbf{S_r^2}(t) = \frac{1}{m-1} \sum_{j=1}^m \frac{w_j^2}{\overline{w}^2} \widetilde{\widetilde{\mathbf{r}}}_j(t) \widetilde{\widetilde{\mathbf{r}}}_j(t)',$$

$$S_\varepsilon^2(1,0) = \frac{1}{m-1} \sum_{j=1}^m \frac{w_j^2}{\overline{w}^2} \widetilde{\widetilde{\varepsilon}}_j(1) \widetilde{\widetilde{\varepsilon}}_j(0), \qquad \mathbf{S_r^2}(1,0) = \frac{1}{m-1} \sum_{j=1}^m \frac{w_j^2}{\overline{w}^2} \widetilde{\widetilde{\mathbf{r}}}_j(1) \widetilde{\widetilde{\mathbf{r}}}_j(0)',$$

$$S^2(w) = \frac{1}{m-1} \sum_{j=1}^m (w_j - \overline{w})^2,$$

$$\mathbf{S_x^2} = \frac{1}{m} \sum_{j=1}^m \frac{w_j}{\overline{w}} \widetilde{\widetilde{\mathbf{x}}}_j' \widetilde{\widetilde{\mathbf{x}}}_j, \quad \mathbf{S_{x,Y}^2}(t) = \frac{1}{m} \sum_{j=1}^m \frac{w_j}{\overline{w}} \widetilde{\widetilde{\mathbf{x}}}_j' \widetilde{\widetilde{Y}}_j(t),$$

and

$$\mathbf{S_{xY}^2}(t) = \frac{1}{m} \sum_{j=1}^m (\frac{w_j}{\overline{w}} \widetilde{\widetilde{\mathbf{x}}}_j' \widetilde{\widetilde{Y}}_j(t) - \overline{wxY}(t))^2,$$

where $\overline{wxY}(t) = \frac{1}{m} \sum_{j=1}^m \frac{w_j}{\overline{w}} \widetilde{\widetilde{\mathbf{x}}}_j' \widetilde{\widetilde{Y}}_j(t)$. Finally, we define several variance terms pertaining to

WLS estimates of the ATE and covariate parameters in the RCT working model:

$$V_{ATE} = \frac{V_\varepsilon}{[p(1-p)]^2} = \frac{(1-p)}{p} S_\varepsilon^2(1) + \frac{p}{(1-p)} S_\varepsilon^2(0) + 2S_\varepsilon^2(1,0)$$

and

$$\mathbf{V_\gamma} = (\mathbf{S_x^2})^{-1} \mathbf{V_r} (\mathbf{S_x^2})^{-1},$$

where $\mathbf{V_r} = p(1-p)[\mathbf{S_r^2}(1) + \mathbf{S_r^2}(0) - 2\mathbf{S_r^2}(1,0)]$.

We can now state our theorem. All specified conditions, apart from (C3), are required for the

CLT, whereas (C3) specifies a condition on the rate of convergence of $\lambda_m$ as $m$ gets large.

**Theorem 1.** Assume SUTVA (Condition (C1)), randomization (C2), and the following 6

additional conditions for $t \in \{1,0\}$:

(C3)   As $m \to \infty$, $\frac{\lambda_m}{\sqrt{m}} \to \lambda_0 \geq 0$.

(C4)   Let $h(t) = \max_{1 \leq j \leq m} \left( \frac{w_j^2}{\overline{w}^2} \widetilde{\widetilde{\varepsilon}}_j(t)^2 \right)$ and as $m \to \infty$,

$$\max_{t \in \{1,0\}} \frac{1}{(m^t)^2} \frac{h(t)}{(V_{ATE}/m)} \to 0.$$

(C5)    Let $g_q(t) = \max_{1 \le j \le m} \left( \frac{w_j^2}{\overline{w}^2} ([\tilde{\tilde{\mathbf{x}}}_j]_q^2) \right)$ for each covariate $q = 1, \dots, k$ and as $m \to \infty$,

$$\frac{1}{\min(m^1, m^0)} \frac{g_q(t)}{[\mathbf{S}_{\mathbf{x}}^2]_{q,q}} \to 0.$$

(C6)    Let $\theta_q(t) = \max_{1 \le j \le m} \left( \frac{w_j^2}{\overline{w}^2} ([\tilde{\tilde{\mathbf{r}}}_j(t)]_q^2) \right)$ for $q = 1, \dots, k$ and as $m \to \infty$,

$$\max_{t \in \{1,0\}} \frac{1}{(m^t)^2} \frac{\theta_q(t)}{([\mathbf{V}_{\mathbf{Y}}]_{q,q}/m)} \to 0.$$

(C7)    $f^t = \frac{m^t}{m}$ has a limiting value between 0 and 1 and as $m \to \infty$,

$$(1 - f^t) \frac{S^2(w)}{m^t \overline{w}^2} \to 0.$$

(C8)    $\mathbf{S}_{\mathbf{r}}^2(t), S_{\varepsilon}^2(t), \mathbf{S}_{\mathbf{r}}^2(1,0), S_{\varepsilon}^2(1,0), \mathbf{S}_{\mathbf{x}}^2, \mathbf{S}_{\mathbf{x},Y}^2(t),$ and $\mathbf{S}_{\mathbf{x}Y}^2(t)$ have finite (positive definite) limiting values.

Then as $m \to \infty$, the lasso estimator, $\widehat{\boldsymbol{\delta}}_{lasso}$, satisfies

$$\sqrt{m}(\widehat{\boldsymbol{\delta}}_{lasso} - \boldsymbol{\delta}) \xrightarrow{d} \operatorname{argmin}_{\mathbf{u}}(V),$$

where the function $V$ is

$$V(\mathbf{u}) = -2\mathbf{u}'\mathbf{W} + \mathbf{u}'\mathbf{Q}\mathbf{u} + \lambda_0 \sum_{q=1}^{k} u_q \operatorname{sgn}(\delta_q) + \lambda_0 \sum_{q=(k+1)}^{v+1} |u_q|, \qquad (6)$$

where $\mathbf{u} = (u_1, \dots, u_{v+1})'$ is a vector of real numbers; $\operatorname{sgn}(.) \in \{-1,0,1\}$ is the sign function;

$$\mathbf{Q} = \begin{bmatrix} p(1-p) & \mathbf{0} \\ \mathbf{0} & \mathbf{\Omega}_{\mathbf{x}}^2 \end{bmatrix};$$

and

$$\mathbf{W} \text{ has a } N\left(\mathbf{0}, \begin{bmatrix} \Omega_\varepsilon^2 & \mathbf{0} \\ \mathbf{0} & \mathbf{\Omega}_{\mathbf{r}}^2 \end{bmatrix}\right) \text{ distribution;}$$

with $\mathbf{\Omega}_{\mathbf{x}}^2 = \lim_{m \to \infty} \mathbf{S}_{\mathbf{x}}^2$; $\Omega_\varepsilon^2 = \lim_{m \to \infty} V_\epsilon$; and $\mathbf{\Omega}_{\mathbf{r}}^2 = \lim_{m \to \infty} \mathbf{V}_{\mathbf{r}}$.

*Remark 1.* Condition (C7) ensures the stability of the treatment assignment probabilities and provides a weak law of large numbers for the weights. Condition (C8) ensures limiting values on asymptotic variances and covariances, while Conditions (C4), (C5), and (C6) are Lindeberg-type conditions required to invoke finite population CLTs using results in Schochet et al.[17] and Li and Ding[46] , where (C6) is added in our context (see Appendix A).

*Remark 2.* If $\lambda_0 = 0$, minimizing $V(\mathbf{u})$ with respect to $\mathbf{u}$ yields $\operatorname{argmin}_{\mathbf{u}}(V) = \mathbf{Q}^{-1}\mathbf{W}$, in which case we have that $\sqrt{m}(\widehat{\boldsymbol{\delta}}_{lasso} - \boldsymbol{\delta}) \xrightarrow{d} N(\mathbf{0}, \begin{bmatrix} \Omega_{ATE}^2 & \mathbf{0} \\ \mathbf{0} & \Omega_{\boldsymbol{\gamma}}^2 \end{bmatrix})$, where $\Omega_{ATE}^2 = \lim_{m \to \infty} V_{ATE}$ and $\Omega_{\boldsymbol{\gamma}}^2 = \lim_{m \to \infty} \mathbf{V}_{\boldsymbol{\gamma}} = (\Omega_{\mathbf{x}}^2)^{-1} \Omega_{\mathbf{r}} (\Omega_{\mathbf{x}}^2)^{-1}$. This is the limiting distribution of the WLS estimator for the ATE and covariate parameters in our RCT working model (see Appendix A).

*Remark 3.* Theorem 1 accounts for the error structure under the design-based model discussed above, including heteroscedasticity across the research groups. Wagener and Dette[47] discuss lasso estimation for general forms of heteroscedasticity for non-clustered data.

### 3.3. *Consistency of lasso variable selection*

In what follows, let $\tilde{\tilde{\mathbf{z}}}_j = \begin{bmatrix} \tilde{T}_j & \tilde{\tilde{\mathbf{x}}}_j \end{bmatrix}$ be the vector of explanatory variables with associated parameter vector $\boldsymbol{\delta} = (\beta_1 \ \boldsymbol{\gamma}')'$. We partition the parameter vector as $\boldsymbol{\delta} = (\boldsymbol{\delta}_I' \ \boldsymbol{\delta}_N')'$, where $\boldsymbol{\delta}_I$ are the $k$ nonzero (included) parameters and $\boldsymbol{\delta}_N = \mathbf{0}$ are the $(v + 1 - k)$ zero (non-included) parameters. We similarly partition the explanatory variables as $\tilde{\tilde{\mathbf{z}}}_j = (\tilde{\tilde{\mathbf{z}}}_{I,j} \ \tilde{\tilde{\mathbf{z}}}_{N,j})$ and the variables in (6) as $\mathbf{u} = \begin{pmatrix} \mathbf{u}_I \\ \mathbf{u}_N \end{pmatrix}$, $\mathbf{W} = \begin{pmatrix} \mathbf{W}_I \\ \mathbf{W}_N \end{pmatrix}$, and $\mathbf{Q} = \begin{bmatrix} \mathbf{Q}_{II} & \mathbf{Q}_{IN} \\ \mathbf{Q}_{NI} & \mathbf{Q}_{NN} \end{bmatrix}$.

For the analysis, we label the estimator $\widehat{\boldsymbol{\delta}}_{lasso}$ for $\boldsymbol{\delta}$ as consistent for variable selection if $\lim_{m \to \infty} P(\widehat{\boldsymbol{\delta}}_{lasso,N,q} = 0) = 1$ for $q > k$, and as consistent for conservative variable selection if $\lim_{m \to \infty} P(\widehat{\boldsymbol{\delta}}_{lasso,N,q} = 0) = c$ for $q > k$ and $0 < c < 1$.[48,49] To examine variable selection consistency in our setting, note that the first-order conditions for minimizing (6) with respect to $\mathbf{u}_I$ and $\mathbf{u}_N$ are as follows:

$$\mathbf{Q}_{II}\hat{\mathbf{u}}_I + \mathbf{Q}_{IN}\hat{\mathbf{u}}_N - \mathbf{W}_I = -\frac{\lambda_0}{2}\begin{pmatrix}\text{sgn}(\delta_1)\\ \vdots \\ \text{sgn}(\delta_k)\end{pmatrix} = -\frac{\lambda_0}{2}s(\boldsymbol{\delta})$$

and

$$|\mathbf{Q}_{NN}\hat{\mathbf{u}}_N + \mathbf{Q}_{NI}\hat{\mathbf{u}}_I - \mathbf{W}_N| \leq \frac{\lambda_0}{2}\mathbf{1}, \tag{7}$$

where $\mathbf{1}$ is a $(v + 1 - k)$ vector of 1s. It follows that $\hat{\mathbf{u}}_N = \mathbf{0}$ if

$$\hat{\mathbf{u}}_I = \mathbf{Q}_{II}^{-1}\left(\mathbf{W}_I - \frac{\lambda_0}{2}s(\boldsymbol{\delta})\right) \sim N(-\frac{\lambda_0}{2}\mathbf{Q}_{II}^{-1}s(\boldsymbol{\delta}), \mathbf{Q}_{II}^{-1}\mathbf{W}_I\mathbf{Q}_{II}^{-1}) \tag{8}$$

and

$$-\frac{\lambda_0}{2}\mathbf{1} \leq \mathbf{Q}_{NI}\mathbf{Q}_{II}^{-1}\left(\mathbf{W}_I - \frac{\lambda_0}{2}s(\boldsymbol{\delta})\right) - \mathbf{W}_N \leq \frac{\lambda_0}{2}\mathbf{1}, \tag{9}$$

where we obtain (9) by plugging the first part of (8) into (7).

The inequality in (9) establishes that if $\lambda_0 > 0$, there is a positive probability that lasso estimates all components of $\boldsymbol{\delta}_N$ as zero, but this probability could be less than 1, in which case lasso performs conservative model selection (if $\lambda_0 = 0$, the probability is zero as implied by *Remark 2* above). Thus, if $\lambda_0 > 0$, the rate of convergence of $(\hat{\boldsymbol{\delta}}_{lasso} - \boldsymbol{\delta})$ will be $\sqrt{m}$, but lasso could select incorrect covariates with positive asymptotic probability without added conditions.

Several articles have developed such added conditions so that lasso can attain consistent lasso variable selection.[20,50,51] An almost necessary and sufficient condition for this to occur is that there exists a $kx1$ positive constant vector, $\boldsymbol{\eta}$, such that $|\mathbf{Q}_{NI}\mathbf{Q}_{II}^{-1}sgn(\boldsymbol{\delta}_I)| \leq \mathbf{1} - \boldsymbol{\eta}$.[50] In this case, the asymptotic variance of the lasso estimator for the non-zero parameters in (8) is the WLS estimator. This "strong irrepressible" condition holds under constrained settings with small (or zero) correlations among the covariates, but violations are common.[50]

## 4. Lasso-OLS Hybrid ATE Estimator

In the second stage of the approach, WLS is used for ATE estimation using the covariates selected by lasso in the first stage, with design-based standard errors estimated using (4), a consistent (upper bound) plug-in estimator for $\frac{1}{m}\Omega^2_{ATE}$. To improve precision, the ATE estimator, $\hat{\beta}_{1,Lasso-OLS}$, is obtained using individual-level data rather than cluster-level averages. Standard errors and asymptotic distributions under this approach do not account for the noise in the lasso covariate selection process itself using recent methods discussed in the literature (this is an area for future extensions).[52,53,54,55] Standard errors also do not account for the potential variation in the selected covariates across different treatment-control allocations (although the use of CV methods can help minimize this problem). Note that this variation is not present under the strong null hypothesis of no intervention effects for any unit, in which case both treatment and control potential outcomes are observed for all units.

## 5. Simulations

We ran simulations to examine the finite sample performance of the two-stage Lasso-OLS hybrid estimator using simulation methods found in the literature. We used the following data generating process to obtain individual-level potential outcomes and covariates under the null hypothesis of zero ATEs:

$$Y_{ij}(0) = \sum_{q=1}^{k} x_{I,ij,q}\gamma_{I,q} + u_j + e_{ij}, \tag{10a}$$

$$Y_{ij}(1) = Y_{ij}(0) + \tau_j + \theta_{ij}, \tag{10b}$$

and

$$x_{ij,q} = u_{Xj} + e_{Xij}, \tag{10c}$$

where $u_j$, $e_{ij}$, $\tau_j$, $\theta_{ij}$, $u_{Xj}$, and $e_{Xij}$ are mean zero random errors drawn independently from normal distributions. To allow for correlated covariates in (10c), the $u_{Xj}$ and $e_{Xij}$ terms were generated from a multivariate normal distribution, $N(0, \Sigma)$, where $\Sigma_{gg} = 1$ or $\Sigma_{gh} = \rho^{|g-h|}$, with $\rho = 0$ or $0.5$ (the irrepressibility condition holds in these cases). The $\gamma_{l,q}$ parameters were generated independently from $t$-distributions with 3 degrees of freedom.[24] To generate the multi-level errors in each equation, we assumed an intraclass correlation coefficient of $0.10$.[4] The variances of $u_j$ and $e_{ij}$ in (10a) were calculated to yield a regression $R^2$ value of $0.5$. The variances of $\tau_j$ and $\theta_{ij}$ in (10b), which capture treatment effect heterogeneity, were obtained assuming their sum is equal to 5 percent of the total variance of $Y_{ij}(0)$.

We conducted simulations for $m = 20, 40$, and $80$ clusters, with $p = 0.6$, where cluster sample sizes ($n_j$) were randomly varied between 40 and 80. The number of included covariates ($k$) was set to $k = 3$ for simulations with $m = 20$ and to $k = 5$ with $m = 40$ or $80$; the total number of covariates ($v$) was set to $v = 10$ or $80$. We generated data once for each combination of simulation parameters and ran 4,000 simulations by sampling from the randomization distribution to assign a treatment status to each cluster. We repeated this process four times to guard against unusual base samples, and then calculated average results. Lasso estimation was conducted using cluster averages, with clusters weighted by their sample sizes; the analysis used the glmselect procedure in the SAS statistical package with leave-one-out cross-validation.

The simulation results are displayed in Table 1. Across replications, lasso selected between 3 and 7 covariates, with larger sets selected for designs with more clusters, more total covariates included in the estimation ($v = 80$ versus $v = 10$), and lower correlations among the covariates ($\rho = 0$ versus $\rho = 0.5$). Lasso correctly selected about 40 to 60 percent of the "true" covariates with non-zero parameters (recall $k = 3$ or $5$), where correct selection rates increased with larger samples and fewer total model covariates.

The simulation results suggest that biases of the two-stage ATE estimates are small, even with relatively small numbers of clusters (Table 1). Type 1 errors hovered between 3 and 5 percent, with values closer to 5 percent for the larger samples. The second stage, design-based standard error estimates in (4) are somewhat conservative, as can be seen by comparing their average values to the "true" values, calculated as the standard deviations of the estimated ATEs from models containing the non-zero covariates only (Table 1) (these findings are similar to those in Bloniarz et al.[24] for non-clustered RCTs). The estimated standard errors become less conservative with more clusters. Across replications, about 95 percent of the estimated confidence intervals contain zero, with confidence intervals calculated as $\hat{\beta}_{1,Lasso-OLS} \pm$

$T^{-1}(1 - \frac{\alpha}{2}, df)\sqrt{\text{V}\hat{a}r\left(\hat{\beta}_{1,Lasso-OLS}\right)}$ using the variance estimator in (4), where $T^{-1}$ is the inverse of the t-distribution with $df$ degrees of freedom and $\alpha = 0.05$ (Table 1).

## 6. Empirical example

To demonstrate the Lasso-OLS hybrid procedure, we used RCT data from the Social and Character Development (SACD) Research Program, a large-scale, school-based RCT co-funded by the Centers for Disease Control and Prevention and the Institute of Education Sciences at the U.S. Department of Education.[56] The SACD study evaluated promising behavioral health interventions in seven large school districts that were designed to promote positive social and character development among elementary school children, with the goal of improving their academic performance. Within each site, half the schools were randomly assigned to a treatment group and half to a control group, yielding of sample of 84 schools (42 treatment and 42 control).

Intervention features included materials and lessons on social skills, behavior management, social and emotional learning, self-control, anger management, and violence prevention.

Our analysis used baseline and outcome data on 4,018 4th graders (2,147 treatments and 1,871 controls).[56] We relied on six primary study outcome scales (Table 2) and 46 available baseline covariates (Appendix Table B.1). Baseline and follow-up data were obtained from child reports administered in the classroom, primary caregiver telephone interviews, and teacher reports on students. The SACD study is a good test case for our estimators because many scale items were tailored and factor-analyzed for the study, making it difficult to fully anticipate outcome-covariate relationships in advance. The actual SACD evaluation did not pre-specify covariates but used stepwise regression without cross-validation to select covariates, and also selected covariates based on their correlations with treatment status.[56] The goal of our analysis is not to replicate evaluation findings, but to demonstrate our ATE estimation approach.

We estimated the lasso model using cluster (school) averages and leave-one-out cross-validation, where clusters were weighted by student sample sizes. We first identified predictive main effects and then predictive two-way interactions among the predictive main effects; all models included treatment status and site indicators. Next, we estimated ATEs and design-based standard errors using the individual data and regression models that included the lasso-selected covariates from the first stage.

Table 3 and Appendix Table B.1 present the lasso results. Table 3 also compares ATE findings from our Lasso-OLS hybrid approach to findings from a commonly used RCT approach where the pre-intervention measure of the primary outcome (the pretest) is used as the model covariate for that outcome. This latter approach yields a pretest-posttest ATE estimator and is a typical pre-specification approach used across fields, when pre-specification does in fact occur.

We find that lasso selected a parsimonious set of covariates using the SACD data—between 2 and 5 covariates across the 6 outcomes (Table 3). At least one pretest was selected for all 6 outcomes, and the specific pretest aligned with the posttest was selected for 4 outcomes (Table B.1). Only one predictive interaction term was identified (for the altruistic scale from the caregiver reports). In addition, several demographic covariates were commonly selected across the 6 outcomes, including education of the household head, the presence of both parents in the household, and household income level (Table B.1).

The results in Table 3 indicate that the ATE point estimates are very similar from models with the lasso-selected covariates and those with the "pre-specified" pretest only. Standard error reductions from using the lasso-selected covariates range from 7 to 22 percent across outcomes, with a mean of 15 percent. While none of the estimated ATEs are statistically significant at the 5 percent level using either approach, the impact on the scale measuring student fear at school is marginally statistically significant at the 10 percent level using the lasso-selected covariates ($p$-value $= 0.067$). This suggests that, due to potential precision gains, our two-stage approach could influence study findings relative to an approach with covariate pre-specification only.

## 7. Conclusions

This article developed a Lasso-OLS hybrid approach for clustered RCTs that involves pre-specifying a process for identifying predictive covariates for ATE estimation once the outcome data have been collected, rather than pre-specifying the covariates themselves. The two-stage procedure is consistent with design-based principles for regression estimators underlying experiments and is simple to apply using existing software. In the first stage, lasso is estimated using cluster-level averages and weights, applying leave-one-out cross-validation to ensure replicability and feasibility for designs with relatively small numbers of clusters. In the second stage, ATEs and design-based standard errors are calculated using WLS run on the individual-level data, where the regression model includes the lasso covariates selected from the first stage.

Our design-based approach applies to continuous, binary, and discrete outcomes and is nonparametric. Further, the lasso estimator is asymptotically normal and can achieve variable selection consistency, generalizing the results of Knight and Fu[18] to clustered RCT setting. These asymptotic results rely on a new central limit theorem for finite population regression estimators for clustered RCTs. The simulation results indicate that Type 1 errors of the ATE estimates are near nominal values, even with relatively small numbers of clusters, and standard errors are close to the true ones, although somewhat conservative with small samples. The empirical analysis demonstrated that the approach can improve precision, without sacrificing technical rigor.

We recommend pre-specification of covariates as the preferred strategy for RCTs. However, there might not be always be a scientific basis for identifying predictive covariates a priori. Further, limited statistical power is often a reality for clustered RCTs due to the expense of adding study clusters (such as hospitals, schools, or communities), so increasing model $R^2$ values may be needed to achieve precision targets. In these cases, the methods discussed in this article are applicable, and can overcome some of the usual criticisms of post-hoc covariate selection in RCT analyses, such as model overfitting and lack of replicability and transparency.

Future extensions could establish the asymptotic properties of the first-stage lasso estimator for the more general case where the number of covariates grows with the number of clusters. Future research could also examine the asymptotic distributions of the second stage ATE estimators that account for the first stage lasso covariate selection process. Finally, the results could be extended to commonly used variants of lasso found in the literature, such as adaptive lasso and the elastic net.[51,57]

**Table 1.** Simulation results using the Lasso-OLS two-stage procedure.

| Parameter values† | Number of covariates selected by lasso | | Average bias, Type 1 error, and average 95% confidence interval coverage of estimated ATEs | | | Standard errors of estimated ATEs††† | |
|---|---|---|---|---|---|---|---|
| | Total (true and not) | True | Average Bias†† | Type 1 error | Coverage rate | Average estimate | True |
| **_m_ = 20, _k_ = 3** | | | | | | | |
| $v = 10$, $\rho = 0.0$ | 2.9 | 1.7 | -0.002 | 0.041 | 0.959 | 0.950 | 0.897 |
| $v = 10$, $\rho = 0.5$ | 2.6 | 1.4 | 0.000 | 0.044 | 0.956 | 0.373 | 0.322 |
| $v = 80$, $\rho = 0.0$ | 4.4 | 1.6 | -0.009 | 0.028 | 0.972 | 0.308 | 0.296 |
| $v = 80$, $\rho = 0.5$ | 3.5 | 0.5 | 0.006 | 0.035 | 0.965 | 0.249 | 0.203 |
| **_m_ = 40, _k_ = 5** | | | | | | | |
| $v = 10$, $\rho = 0.0$ | 3.9 | 2.9 | 0.000 | 0.047 | 0.953 | 0.257 | 0.236 |
| $v = 10$, $\rho = 0.5$ | 3.0 | 2.3 | -0.001 | 0.043 | 0.957 | 0.489 | 0.464 |
| $v = 80$, $\rho = 0.0$ | 6.2 | 2.1 | 0.000 | 0.041 | 0.959 | 0.474 | 0.451 |
| $v = 80$, $\rho = 0.5$ | 3.5 | 1.9 | 0.001 | 0.042 | 0.959 | 0.464 | 0.412 |
| **_m_ = 80, _k_ = 5** | | | | | | | |
| $v = 10$, $\rho = 0.0$ | 4.6 | 2.8 | 0.001 | 0.045 | 0.955 | 0.384 | 0.376 |
| $v = 10$, $\rho = 0.5$ | 4.3 | 2.9 | 0.000 | 0.050 | 0.950 | 0.283 | 0.274 |
| $v = 80$, $\rho = 0.0$ | 7.2 | 3.3 | 0.000 | 0.045 | 0.955 | 0.283 | 0.277 |
| $v = 80$, $\rho = 0.5$ | 6.4 | 3.0 | -0.001 | 0.042 | 0.958 | 0.128 | 0.120 |

† The parameter, $m$, is the total number of clusters; $k$ is the number of (true) covariates with non-zero parameters; $v$ is the total number of covariates (both zero and non-zero parameters); and $\rho$ is the correlation between adjacent covariates. See the text for simulation assumptions and methods.

††The bias is calculated relative to the estimated ATEs from the true model.

†††The "average estimate" is the average standard error across replications using the design-based standard error estimator in (4) from models with the first stage lasso covariates. The "true" standard error is the standard deviation of ATE estimates across replications from the model with the true covariates.

**Table 2.** Outcome variables for the empirical analysis using SACD RCT data.

| Outcome | Data source (Spring 2007) | Description of variable |
|---|---|---|
| Problem behavior | Child report | Scale ranges from 0 to 3 and contains 6 items from the Frequency of Delinquent Behavior Scale and 6 items from the Aggression Scale; Reliability = 0.86. |
| Normative beliefs about aggression | Child report | Scale ranges from 1 to 4 and contains 12 items from the Normative Beliefs About Aggression Scale; Reliability = 0.83. |
| Student afraid at school | Child report | Scale ranges from 1 to 4 and contains 4 items from the Feelings of Safety at School scale; Reliability = 0.79. |
| Altruistic behavior | Primary caregiver report | Scale ranges from 1 to 4 and contains 8 items from the Altruism Scale, Primary Caregiver Version; Reliability = 0.88 |
| Problem behavior | Teacher report | Scale ranges from 1 to 4 and contains 14 items from the BASC Aggression Subscale, Teacher Version, 7 items from the BASC Conduct Problems Subscale, Teacher Version and 2 items from the Responsibility Scale; Reliability = 0.95. |
| Positive social behavior | Teacher report | Scale ranges from 1 to 4 and contains 6 items from the Responsibility Scale and 19 items from the Social Competence Scale and 8 items from the Altruism Scale, Teacher Version; Reliability = 0.97. |

Note: See SACD Research Consortium[56] for a complete description of the construction of these scales.

**Table 3.** Results for the empirical analysis using SACD RCT data.

| Outcome scale | Number of lasso-selected covariates | Second stage ATE estimates | | Second stage standard errors (p-values in parentheses) | |
|---|---|---|---|---|---|
| | | Lasso-selected covariates | Pretest covariate only | Lasso-selected covariates | Pretest covariate only |
| Problem behavior (CR) | 2 | -0.006 | -0.004 | 0.026 (0.801) | 0.028 (0.902) |
| Normative beliefs about aggression (CR) | 3 | -0.005 | 0.003 | 0.025 (0.842) | 0.030 (0.912) |
| Student afraid at school (CR) | 4 | -0.067 | -0.065 | 0.036 (0.067) | 0.044 (0.141) |
| Altruistic behavior (PCR) | 5 | -0.017 | -0.016 | 0.026 (0.521) | 0.028 (0.579) |
| Problem behavior (TR) | 4 | -0.009 | -0.002 | 0.024 (0.695) | 0.028 (0.936) |
| Positive social behavior (TR) | 5 | -0.010 | -0.031 | 0.042 (0.804) | 0.053 (0.561) |

Abbreviations. CR = child report, PCR = primary caregiver report, TR = teacher report.

# Appendix A. Proof of Theorem 1

To prove Theorem 1, we first need to prove a finite population CLT for our design-based working regression model that extends results in Schochet et al.[17] to focus not only on ATE estimation but also on covariate estimation.

**Theorem A.1.** Consider WLS estimation of the working model in (3) using cluster-level averages and weights, $w_j$, where centered cluster-level outcomes, $\tilde{\bar{y}}_j = \bar{y}_j - \bar{\bar{y}}$, are regressed on the centered cluster-level vector, $\tilde{\tilde{\mathbf{z}}}_j = \begin{bmatrix} \tilde{T}_j & \tilde{\mathbf{x}}_j \end{bmatrix}$, with associated parameter vector $\boldsymbol{\delta} = (\beta_1 \ \boldsymbol{\gamma}')'$. Assume also Conditions (C1), (C2), and (C4)-(C8) in Theorem 1 in the main text. Then as $m \to \infty$, we have that

$$\sqrt{m}(\widehat{\boldsymbol{\delta}} - \boldsymbol{\delta}) = \sqrt{m}[\begin{pmatrix} \hat{\beta}_1 \\ \widehat{\boldsymbol{\gamma}} \end{pmatrix} - \begin{pmatrix} \beta_1 \\ \boldsymbol{\gamma} \end{pmatrix}] \xrightarrow{d} N\left(\mathbf{0}, \begin{bmatrix} \Omega_{ATE}^2 & \mathbf{0} \\ \mathbf{0} & \Omega_{\boldsymbol{\gamma}}^2 \end{bmatrix}\right), \tag{A.1}$$

where $\boldsymbol{\gamma} = (\Omega_{\mathbf{x}}^2)^{-1}\left(p\Omega_{\mathbf{x},Y}^2(1) + (1-p)\Omega_{\mathbf{x},Y}^2(0)\right)$ is the asymptotic covariate parameter vector; $\Omega_{\mathbf{x},Y}^2(t)$ is the limiting value of $\mathbf{S}_{\mathbf{x},Y}^2(t)$ for $t \in \{1,0\}$; $\Omega_{ATE}^2 = \lim_{m \to \infty} V_{ATE} = \frac{1}{[p(1-p)]^2}\Omega_\varepsilon$; and $\Omega_{\boldsymbol{\gamma}}^2 = \lim_{m \to \infty} \mathbf{V}_{\boldsymbol{\gamma}} = (\Omega_{\mathbf{x}}^2)^{-1}\Omega_{\mathbf{r}}(\Omega_{\mathbf{x}}^2)^{-1}$.

*Proof of Theorem A.1*

Schochet et al.[17] prove that under the design-based working model, $\hat{\beta}_1$ is consistent and asymptotically normal with variance $\frac{1}{m}\Omega_{ATE}^2$. They also prove that $\widehat{\boldsymbol{\gamma}} \xrightarrow{p} \boldsymbol{\gamma}$. Thus, we do not repeat those proofs here. To prove the asymptotic normality of $\widehat{\boldsymbol{\gamma}}$, we first write the estimated parameter vector as

$$\begin{pmatrix} \hat{\beta}_1 \\ \widehat{\boldsymbol{\gamma}} \end{pmatrix} = \left(\sum_{j=1}^m w_j \tilde{\tilde{\mathbf{z}}}_j' \tilde{\tilde{\mathbf{z}}}_j\right)^{-1} \left(\sum_{j=1}^m w_j \tilde{\tilde{\mathbf{z}}}_j' \tilde{\bar{y}}_j\right),$$

so that

$$\sqrt{m}[\begin{pmatrix}\hat{\beta}_1\\ \hat{\boldsymbol{\gamma}}\end{pmatrix} - \begin{pmatrix}\beta_1\\ \boldsymbol{\gamma}\end{pmatrix}] = \left(\frac{1}{m}\sum_{j=1}^{m} w_j \tilde{\mathbf{z}}_j' \tilde{\mathbf{z}}_j\right)^{-1} \left(\frac{1}{\sqrt{m}}\sum_{j=1}^{m} w_j \tilde{\mathbf{z}}_j' \tilde{\tilde{l}}_j\right),$$

where $\tilde{\tilde{l}}_j = \tilde{\tilde{y}}_j - \tilde{T}_j \beta_1 - \tilde{\tilde{\mathbf{x}}}_j \boldsymbol{\gamma}$ is the cluster-level residual. Multiplying this out, we find that

$$\sqrt{m}\left[\begin{pmatrix}\hat{\beta}_1\\ \hat{\boldsymbol{\gamma}}\end{pmatrix} - \begin{pmatrix}\beta_1\\ \boldsymbol{\gamma}\end{pmatrix}\right] = \begin{bmatrix} \dfrac{1}{m}\sum_{j=1}^{m}\dfrac{w_j}{\bar{w}}\tilde{T}_j^2 & \dfrac{1}{m}\sum_{j=1}^{m}\dfrac{w_j}{\bar{w}}\tilde{T}_j\,\tilde{\tilde{\mathbf{x}}}_j \\ \dfrac{1}{m}\sum_{j=1}^{m}\dfrac{w_j}{\bar{w}}\tilde{\tilde{\mathbf{x}}}_j'\tilde{T}_j & \dfrac{1}{m}\sum_{j=1}^{m}\dfrac{w_j}{\bar{w}}\tilde{\tilde{\mathbf{x}}}_j'\tilde{\tilde{\mathbf{x}}}_j \end{bmatrix}^{-1} \begin{bmatrix} \dfrac{1}{\sqrt{m}}\sum_{j=1}^{m}\dfrac{w_j}{\bar{w}}\tilde{T}_j\tilde{\tilde{l}}_j \\ \dfrac{1}{\sqrt{m}}\sum_{j=1}^{m}\dfrac{w_j}{\bar{w}}\tilde{\tilde{\mathbf{x}}}_j'\tilde{\tilde{l}}_j \end{bmatrix},$$

which, to simplify notation, we express as

$$\begin{bmatrix} A & \mathbf{B} \\ \mathbf{B}' & \mathbf{C} \end{bmatrix}^{-1} \begin{bmatrix} E \\ \mathbf{F} \end{bmatrix}. \tag{A.2}$$

Applying standard results on inverses of partitioned matrices to (A.2), we find after some algebra that

$$\sqrt{m}(\hat{\boldsymbol{\gamma}} - \boldsymbol{\gamma}) = \frac{-\mathbf{C}^{-1}\mathbf{B}'}{Denom}E + \left(\mathbf{C}^{-1} + \frac{\mathbf{C}^{-1}\mathbf{B}'\mathbf{B}\mathbf{C}^{-1}}{Denom}\right)\mathbf{F}, \tag{A.3}$$

where $Denom = (A - \mathbf{B}\mathbf{C}^{-1}\mathbf{B}')$. We now consider each term in (A.3). Under Condition (C7) in the main text, we have $A \xrightarrow{p} p(1-p)$, and because of randomization, we have $\mathbf{B} \xrightarrow{p} \mathbf{0}$, which also implies that $\hat{\beta}_1$ and $\hat{\boldsymbol{\gamma}}$ are asymptotically uncorrelated. Further, using (C8) and Slutsky's theorem, we have $\mathbf{C}^{-1} \xrightarrow{p} (\boldsymbol{\Omega}_{\mathbf{x}}^2)^{-1}$. These results establish that $Denom \xrightarrow{p} A$, $\mathbf{C}^{-1}\mathbf{B}' \xrightarrow{p} \mathbf{0}$, and $\mathbf{C}^{-1}\mathbf{B}'\mathbf{B}\mathbf{C}^{-1}$ $\xrightarrow{p} \mathbf{0}$. Next, under (C1), (C2), (C4), (C5), (C7), and (C8), Schochet et al.[17] prove that $E$ $\xrightarrow{d} N(0, \Omega_\varepsilon^2)$. Thus, the first term on the right-hand side in (A.3) vanishes as $m$ gets large, as does the second term inside the brackets. What remains is to establish a CLT for $\mathbf{F}$.

To do this, it is useful to express $E$ and $\mathbf{F}$ in a parallel form, where we insert the identities, $\beta_1 = \bar{\bar{Y}}(1) - \bar{\bar{Y}}(0)$ and $y_{ij} = T_j Y_{ij}(1) + (1 - T_j)Y_{ij}(0)$, into the expression for $\tilde{\tilde{l}}_j$ to yield:

$$E = \frac{1}{\sqrt{m}} \sum_{j=1}^{m} \frac{w_j}{\overline{w}} \left[ T_j (1-p) \widetilde{\varepsilon_j}(1) - (1-T_j) p \widetilde{\varepsilon_j}(0) \right]$$

and

$$\mathbf{F} = \frac{1}{\sqrt{m}} \sum_{j=1}^{m} \frac{w_j}{\overline{w}} \left[ T_j \widetilde{\mathbf{x}}_j' \widetilde{\varepsilon_j}(1) + (1-T_j) \widetilde{\mathbf{x}}_j' \widetilde{\varepsilon_j}(0) \right] = \frac{1}{\sqrt{m}} \sum_{j=1}^{m} \frac{w_j}{\overline{w}} \left[ T_j \widetilde{\mathbf{r}}_j(1) + (1-T_j) \widetilde{\mathbf{r}}_j(0) \right].$$

Using this formulation, the delta method for finite populations[58] can be used to establish a finite population CLT for each covariate in $\mathbf{F}$ using Condition (C6) on $\widetilde{\widetilde{\mathbf{r}}}_j(t)$, that parallels and supplements (C4) on $\widetilde{\varepsilon_j}(t)$ and (C5) on $\widetilde{\widetilde{\mathbf{x}}}_j$ that are required to establish a CLT for the ATE estimator, $\hat{\beta}_1$. Multivariate normality follows because the CLT holds for any linear combination of the covariates under our rank assumption for $\mathbf{\Omega}_{\mathbf{x}}^2$.

The asymptotic variance of $\hat{\boldsymbol{\gamma}}$ can be derived using the delta method for finite populations[58]. Instead, we find it more transparent to use the following standard asymptotic expansion for $\hat{\boldsymbol{\gamma}}$:

$$\sqrt{m}(\hat{\boldsymbol{\gamma}} - \boldsymbol{\gamma}) = (\mathbf{\Omega}_{\mathbf{x}}^2)^{-1} \frac{1}{\sqrt{m}} \sum_{j=1}^{m} \frac{w_j}{\overline{w}} \left[ T_j \widetilde{\mathbf{r}}_j(1) + (1-T_j) \widetilde{\mathbf{r}}_j(0) \right] + o_p(1),$$

where $o_p(1)$ signifies a term that converges in probability to zero. We can then use this expression to calculate $\lim_{m \to \infty} E_R(\hat{\boldsymbol{\gamma}} - \boldsymbol{\gamma})(\hat{\boldsymbol{\gamma}} - \boldsymbol{\gamma})'$, where the expectation is taken with respect to the randomization distribution, $R$. This yields the following asymptotic variance expression:

$$AsyVar(\hat{\boldsymbol{\gamma}}) = \frac{1}{m} (\mathbf{\Omega}_{\mathbf{x}}^2)^{-1} \mathbf{D_r} (\mathbf{\Omega}_{\mathbf{x}}^2)^{-1}, \tag{A.4}$$

where

$$\mathbf{D_r} = p(1-p) \lim_{m \to \infty} \frac{1}{m} \left( \sum_{j=1}^{m} \frac{w_j^2}{\bar{w}^2} \tilde{\bar{\mathbf{r}}}_j(1) \tilde{\bar{\mathbf{r}}}_j(1)' - \frac{1}{(m-1)} \sum_{j=1}^{m} \sum_{h \neq j}^{m} \frac{w_j^2}{\bar{w}^2} \tilde{\bar{\mathbf{r}}}_j(1) \tilde{\bar{\mathbf{r}}}_h(1)' \right.$$

$$+ \sum_{j=1}^{m} \frac{w_j^2}{\bar{w}^2} \tilde{\bar{\mathbf{r}}}_j(0) \tilde{\bar{\mathbf{r}}}_j(0)' - \frac{1}{(m-1)} \sum_{j=1}^{m} \sum_{h \neq j}^{m} \frac{w_j^2}{\bar{w}^2} \tilde{\bar{\mathbf{r}}}_j(0) \tilde{\bar{\mathbf{r}}}_h(0)' \qquad \text{(A.5)}$$

$$\left. - 2 \sum_{j=1}^{m} \frac{w_j^2}{\bar{w}^2} \tilde{\bar{\mathbf{r}}}_j(1) \tilde{\bar{\mathbf{r}}}_j(0)' + 2 \frac{1}{(m-1)} \sum_{j=1}^{m} \sum_{h \neq j}^{m} \frac{w_j^2}{\bar{w}^2} \tilde{\bar{\mathbf{r}}}_j(1) \tilde{\bar{\mathbf{r}}}_h(0)' \right).$$

This result uses the relations, $Var_R(T_j) = p(1-p)$ and $Cov_R(T_j T_q) = -\frac{p(1-p)}{(m-1)}$. Note next that $\sum_{j=1}^{m} w_j \tilde{\bar{\mathbf{r}}}_j(t) = \mathbf{0}$, so it follows that $(\sum_{j=1}^{m} w_j \tilde{\bar{\mathbf{r}}}_j(t))(\sum_{j=1}^{m} w_j \tilde{\bar{\mathbf{r}}}_j(t))' = \mathbf{0}$, and thus, we find that $\sum_j \sum_{h \neq j} w_j^2 \tilde{\bar{\mathbf{r}}}_j(t) \tilde{\bar{\mathbf{r}}}_h(t)' = -\sum_j w_j^2 \tilde{\bar{\mathbf{r}}}_j(t) \tilde{\bar{\mathbf{r}}}_j(t)'$. Similarly, we have that $\sum_j \sum_{h \neq j} w_j^2 \tilde{\bar{\mathbf{r}}}_j(1) \tilde{\bar{\mathbf{r}}}_h(0)' = -\sum_j w_j^2 \tilde{\bar{\mathbf{r}}}_j(1) \tilde{\bar{\mathbf{r}}}_j(0)'$. Plugging these relations into (A.5) and taking limits under assumption (C8) yields

$$\mathbf{D_r} = \mathbf{\Omega_r^2} = p(1-p)[\mathbf{\Omega_r^2}(1) + \mathbf{\Omega_r^2}(0) - 2\mathbf{\Omega_r^2}(1,0)]$$

and

$$AsyVar(\hat{\boldsymbol{\gamma}}) = \frac{1}{m} (\mathbf{\Omega_x^2})^{-1} \mathbf{\Omega_r^2} (\mathbf{\Omega_x^2})^{-1},$$

where $\mathbf{\Omega_r^2}(1), \mathbf{\Omega_r^2}(0)$ and $\mathbf{\Omega_r^2}(1,0)$ are limiting values of $\mathbf{S_r^2}(1), \mathbf{S_r^2}(0)$ and $\mathbf{S_r^2}(1,0)$, respectively. Combining results establishes the asymptotic result in (A.1), which we now use in the proof for Theorem 1.

**Proof of Theorem 1**

Our proof closely follows Knight and Fu[18] adapted to our clustered RCT setting. Let $\tilde{\bar{l}}_j = (\tilde{\bar{y}}_j - \tilde{\bar{\mathbf{z}}}_j \boldsymbol{\delta})$ be the residual from the WLS regression model, and define $V_m(\mathbf{u})$ as follows:

$$V_m(\mathbf{u}) = \sum_{j=1}^{m} \frac{w_j}{\overline{w}} [(\tilde{\tilde{l}}_j - \frac{\mathbf{u}'\tilde{\tilde{\mathbf{z}}}_j}{\sqrt{m}})^2 - \tilde{\tilde{l}}_j^2] \; + \frac{\lambda_m}{\sqrt{m}} \sum_{q=1}^{v+1} \sqrt{m} \left[ \left| \delta_q + \frac{u_q}{\sqrt{m}} \right| - |\delta_q| \right]. \tag{A.6}$$

Note that $V_m(\mathbf{u})$ is convex and is minimized at $\hat{\mathbf{u}} = \sqrt{m}(\hat{\boldsymbol{\delta}}_{lasso} - \boldsymbol{\delta})$, which can be seen by

comparing first order conditions for (A.6) and for the lasso model in (5) in the main text. To

consider the asymptotic properties of $V_m(\mathbf{u})$, we examine each term on the right-hand side of

(A.6). If we multiply out the first term and use Theorem 1 and Slutsky's theorem, we find that for every $\mathbf{u} \in \mathbb{R}^{v+1}$,

$$\sum_{j=1}^{m} \frac{w_j}{\overline{w}} [(\tilde{\tilde{l}}_j - \frac{\mathbf{u}'\tilde{\tilde{\mathbf{z}}}_j}{\sqrt{m}})^2 - \tilde{\tilde{l}}_j^2] = -2\mathbf{u}' \sum_{j=1}^{m} \frac{w_j}{\overline{w}} \frac{\tilde{\tilde{\mathbf{z}}}_j \tilde{\tilde{l}}_j}{\sqrt{m}} + \mathbf{u}'(\sum_{j=1}^{m} \frac{w_j}{\overline{w}} \frac{\tilde{\tilde{\mathbf{z}}}_j' \tilde{\tilde{\mathbf{z}}}_j}{m}) \mathbf{u}$$

$$\xrightarrow{d} -2\mathbf{u}'\mathbf{W} + \mathbf{u}'\mathbf{Q}\mathbf{u},$$

where $\mathbf{W}$ and $\mathbf{Q}$ are defined in Theorem 1 in the main text.

For the second term in (A.6), note that if $\delta_q \neq 0$, then $\sqrt{m} \left[ \left| \delta_q + \frac{u_q}{\sqrt{m}} \right| - |\delta_q| \right] \to u_q \mathrm{sgn}(\delta_q)$,

whereas if $\delta_q = 0$, then $\sqrt{m} \left[ \left| \delta_q + \frac{u_q}{\sqrt{m}} \right| - |\delta_q| \right] = |u_q|$. Further, $\frac{\lambda_m}{\sqrt{m}} \to \lambda_0$ by Condition (C3).

Thus, again using Slutsky's theorem, we find that

$$\frac{\lambda_m}{\sqrt{m}} \sum_{q=1}^{v+1} \sqrt{m} \left[ \left| \delta_q + \frac{u_q}{\sqrt{m}} \right| - |\delta_q| \right] \to \lambda_0 \sum_{q=1}^{v+1} [u_q \mathrm{sgn}(\delta_q) \, I(\delta_q \neq 0) + |u_q| I(\delta_q = 0)],$$

where $I(arg)$ is the indicator function that equals 1 if $arg$ holds and 0 otherwise.

These results establish that $V_m(\mathbf{u}) \xrightarrow{d} V(\mathbf{u})$ as defined in (6) in the main text. Because $V(\mathbf{u})$

has a unique minimum, results in Geyer[59] establish that

$$\mathrm{argmin}_{\mathbf{u}}(V_m) = \sqrt{m}(\hat{\boldsymbol{\delta}}_{lasso} - \boldsymbol{\delta}) \xrightarrow{d} \mathrm{argmin}_{\mathbf{u}}(V),$$

which proves the theorem.

**Appendix Table B.1.** List of all and lasso-selected covariates for the empirical analysis.

| Child-reported covariates | Caregiver-reported covariates | Teacher-reported covariates |
|---|---|---|
| <u>Demographics</u> | <u>Demographics</u> | <u>Child scales</u> |
| Female | Completed high school or equivalent (**5**, **6**) | Academic competence and motivation (**3**) |
| White (non-Hispanic) | Some college | ADHD symptomology |
| Hispanic (Nonwhite) | Bachelor's or higher degree | Altruistic behavior |
| African American (**5**, **6**) | Highest level of education in household: completed high school or equivalent (**3**, **4**) | Positive social behavior |
| Other ethnicity | Highest level of education in household: some college | Problem behavior (**5**, **6**) |
| <u>Child scales</u> | Highest level of education in household: bachelor's or higher degree (**6**) | Parent-teacher involvement |
| Student afraid at school (**3**) | Mother present in home life | Parent involvement |
| Altruistic behavior | Mother and father present (**1**, **5**, **6**) | |
| Empathy | Respondent other than mother or father | |
| Engagement with learning | Number of people in household | |
| Negative school orientation (**2**, **4**[*]) | Household income: $20,000 to $40,000 | |
| Normative beliefs about aggression | Household income: $40,000 to $60,000 | |
| Sense of school as a community | Household income: more than $60,000 | |
| Problem behaviors (**1**, **2**, **4**[*]) | Income-to-poverty ratio: below 135 percent | |
| Self-efficacy for peer interactions | Income-to-poverty ratio: between 135-185 percent | |
| Victimization at school | Income-to-poverty ratio: above 185 percent (**2**, **3**) | |
| | Full-time employment | |
| | Part-time employment | |
| | <u>Parental scales</u> | |
| | Positive parenting subscale | |
| | Confusion, hubbub, and disorder | |
| | Community risk | |
| | <u>Child scales</u> | |
| | Altruistic behavior (**4**) | |
| | Positive social behavior | |
| | Problem behavior | |

Numbers in parentheses indicate the lasso-selected covariates for each outcome scale listed in Table 2 of the main text: **1** = problem behavior (child report), **2** = normative beliefs about aggression (child report), **3** = student afraid at school (child report), **4** = altruistic behavior (caregiver report), **5** = problem behavior (teacher report), and **6** = positive social behavior (teacher report). The **\*** symbol indicates a two-way interaction term selected by lasso.